\documentclass{article}
\usepackage{dcolumn}% Align table columns on decimal point
\usepackage{bm}% bold math
\usepackage{verbatim}       % Defines \begin{comment}, \end{comment}

\usepackage[dvips]{graphicx}
\usepackage{amsthm}
\usepackage{amssymb}
\usepackage{bm}
\usepackage{amstext}
\usepackage{psfrag}
\usepackage{amsmath}
\usepackage{amsfonts}

\newfont{\bb}{msbm10 at 12pt}

\newcommand{\dd}{{\rm d}}
\newcommand{\bd}{\begin{definition}}                %inizia definizione
\newcommand{\ed}{\end{definition}}                  %fine definizione
\newcommand{\bc}{\begin{corollary}}                 %inizia corollario
\newcommand{\ec}{\end{corollary}}                   %fine corollario
\newcommand{\bl}{\begin{lemma}}                     %inizia lemma
\newcommand{\el}{\end{lemma}}                       %fine lemma
\newcommand{\bp}{\begin{proposition}}            %inizia proposizione
\newcommand{\ep}{\end{proposition}}                %fine proposizione
\newcommand{\bere}{\begin{remark}}                  %inizia osservazione
\newcommand{\ere}{\end{remark}}                     %fine oservazione

\newcommand{\bt}{\begin{theorem}}
\newcommand{\et}{\end{theorem}}

\newcommand{\be}{\begin{equation}}
\newcommand{\ee}{\end{equation}}

\newcommand{\bit}{\begin{itemize}}
\newcommand{\eit}{\end{itemize}}
\newtheorem{theorem}{Theorem}[section]
\newtheorem{corollary}[theorem]{Corollary}
\newtheorem{lemma}[theorem]{Lemma}
\newtheorem{proposition}[theorem]{Proposition}
\theoremstyle{definition}
\newtheorem{definition}[theorem]{Definition}
\theoremstyle{remark}
\newtheorem{remark}[theorem]{Remark}
\newtheorem{example}[theorem]{Example}

\begin{document}
%
%\DeclareGraphicsExtensions{.pdf}

\title{ On the causal properties of warped product spacetimes}

\author{E. Minguzzi \footnote{Department of Applied Mathematics, Florence
 University, Via S. Marta 3,  50139 Florence, Italy. E-mail: ettore.minguzzi@unifi.it}}

\date{}
\maketitle

\begin{abstract}
\noindent  It is shown that the warped product spacetime
$P=M\times_f H$, where $H$ is a complete Riemannian manifold, and
the original spacetime $M$ share necessarily the same causality
properties, the only exceptions being the properties of causal
continuity and causal simplicity which present some subtleties. For
instance,   it is shown  that if $\textrm{diam}\,H=+\infty$, the
direct product spacetime $P=M\times H$ is causally simple if and
only if $(M,g)$ is causally simple, the Lorentzian distance on $M$
is continuous and any two causally related events at finite distance
are connected by a maximizing geodesic.  Similar conditions are
found for the causal continuity property. Some new results
concerning the behavior of the Lorentzian distance on
distinguishing, causally continuous, and causally simple spacetimes
are obtained. Finally, a formula which gives the Lorentzian distance
on the direct product in terms of the distances on the two factors
$(M,g)$ and $(H,h)$ is obtained.
\end{abstract}

%\pacs{}

%\noindent Key Words:

\section{Introduction}
At the top of the causal ladder  of spacetimes \cite{minguzzi06c}
\cite{hawking74,beem96,senovilla97} stands the property of {\em
global hyperbolicity} which implies many good properties for the
Lorentzian distance function.
%(time
%separation) defined by
%\[
% d(x,z)=\sup_{\sigma
%\in \mathcal{C}^1(x,z)} l(\sigma),
%\]
% if $z \in J^{+}(x)$ and
%by $d(x,z)=0$ if $z \notin J^{+}(x)$. Here $\mathcal{C}^{1}(x,z)$ is
%the set of (piecewise) $C^1$ future directed causal curves
%connecting $x$ to $z$, and $l(\sigma)=\int_{\sigma} \dd s$ is the
%Lorentzian length of the curve.
Indeed, in a globally hyperbolic spacetime $(M,g)$ the Lorentzian
distance $d: M\times M \to[0,+\infty]$ is
\begin{itemize}
\item[(a)] finite \cite[Lemma 4.5]{beem96},
\item[(b)] maximized by a suitable connecting causal geodesic $\sigma$,
$l(\sigma)=d(x,z)$, whenever $z \in J^{+}(x)$, $d(x,z)<+\infty$,
(Avez-Seifert theorem \cite{avez63,seifert67}) \cite[Theor. 3.18,
Prop. 10.39]{beem96}, \cite[Prop. 14.19]{oneill83}, \cite[Prop.
6.7.1]{hawking73},
\item[(c)] continuous \cite[Lemma 4.5]{beem96}.
\end{itemize}
All these properties are lost even in spacetimes  sharing  the
causal property which stays immediately below global hyperbolicity
in the causal ladder, i.e. {\em causal simplicity}. In particular
even if $d(x,z)<+\infty$ there can be no connecting maximizing
geodesic (see figure \ref{fig1} and  figure 10 in
\cite{minguzzi06c}).

Apparently unrelated with the previous aspects of the Lorentzian
distance function in causally simple spacetimes stands the problem
of finding under which conditions the warped product spacetime $P=M
\times_f H$, endowed with the warped metric $\hat{g}=g+f^2 h$, is
causally simple. For other causal properties such as being
chronological, causal, strongly causal, stably causal or globally
hyperbolic, it has been proved \cite[Prop.
3.61,3.62,3.64,3.68]{beem96}  that they are shared by $(P,\hat{g})$
if and only if they are shared by $(M,g)$. I shall prove that the
same holds for non-total viciousness and the distinction property,
but for causal continuity and causal simplicity this simple
correspondence does not hold. For instance for causal simplicity in
the simple case $f=1$, $(H,h)=(\mathbb{R},\dd y^2)$, one has to
require that $(M,g)$ satisfies also properties (b) and (c) above.

I refer the reader to \cite{minguzzi06c} for most of the conventions
used in this work. In particular, I denote with $(M,g)$ a $C^{r}$
spacetime (connected, time-oriented Lorentzian manifold), $r\in \{2,
\dots, \infty\}$ of arbitrary dimension $n\geq 2$ and signature
$(-,+,\dots,+)$. On $M\times M$ the usual product topology is
defined.  With $\bm{g}$ it is denoted the class of metrics conformal
to $g$. With $(M,\bm{g})$ it is denoted the conformal structure i.e.
the class of spacetimes $[(M,g)]$ on the same manifold $M$, with
metrics related by a conformal rescaling, and the same time
orientation. Sometimes I write ``spacetime $(M,\bm{g})$'' although
by this I mean the conformal structure. With ``lightlike geodesic
$\gamma$ of $(M,\bm{g})$'' I mean a curve which is a lightlike
pregeodesic for a representative $g \in \bm{g}$ and hence for every
element of the class $\bm{g}$.
%In general the mention to the parametrization is omitted since it is
%defined up to reparametrization which preserve the time orientation.

For convenience and generality I often use the causal relations on
$M \times M$ in place of the more widespread point based relations
$I^{+}(x)$, $J^{+}(x)$, $E^{+}(x)$ (and past versions). Recall
\cite{minguzzi06c} the following definition of sets on $M \times M$
\[
I^{+}=\{(p,q) : p\ll q  \} , \quad J^{+}=\{(p,q) : p\leq q  \},
\quad E^{+}=\{(p,q) : p\rightarrow q  \}.
\]
Clearly, $E^+={J^+}\backslash I^+$. Moreover, $I^{+}$ is open
\cite[Chap. 14, Lemma 3]{oneill83} \cite[Prop. 2.16]{minguzzi06c},
$\bar{J}^+=\bar{I}^+$, $\textrm{Int}\, J^+=I^+$ and
$\dot{J}^+=\dot{I}^+$ \cite[Prop. 2.17]{minguzzi06c}.  Finally,
$J^{+}$ is closed in causally simple spacetimes \cite[Prop.
3.68]{minguzzi06c}.  The {\em vanishing  distance set} is the set $(I^+)^C$  which is  made  of all the pairs of events at which the Lorentzian distance $d$ vanishes  (this set is a conformal invariant concept  although the Lorentzian distance is not). Recall that $d$ is lower semi-continuous \cite[Lemma 4.4]{beem96}, in particular it is continuous at those points where $d(x,z)=+\infty$ and in the open set $(\bar{I}^+)^C$, where it vanishes identically.

Most of the causal vectors and curves that we shall encounter will
be {\em future directed}, thus, for simplicity, I  shall omit this
adjective.

\section{The spacetime $P=M\times H$ and warped products}

Consider the spacetime $(P, \tilde{g})$, $P=M \times H$,
$\tilde{g}=g+ h$, where $(H,h)$ is a complete Riemannian manifold
and let $\rho: H\times H \to [0,+\infty)$ be the Riemannian distance
(the assumption of completeness for $(H,h)$ is needed only for a
subset of the results proved below). The length of a causal curve on
$M$ will be denoted with $l$ while the length of a curve on $H$ will
be denoted with $l_H$. The Riemannian distance $\rho$ is continuous
\cite[p. 3]{beem96} while the Lorentzian distance $d: M\to
[0,+\infty]$ is only lower semi-continuous in general. By the
Hopf-Rinow (Heine-Borel) theorem any closed and bounded subset of
the complete Riemannian manifold $H$ is compact, in particular if
$\textrm{diam}(H,h)=\sup_{y_0,y_1 \in H} \rho(y_0,y_1)\, <+\infty$
then $H$ is compact and since $\rho$ is continuous it attains the
supremum, that is, there are $y_0,y_1 \in H$ such that
$\rho(y_0,y_1)=\textrm{diam}(H,h)$.

One of the simplest choices for $(H,h)$ is $(\mathbb{R},\dd y^{2})$,
$y \in \mathbb{R}$, in which case $\rho(y_0,y_1)=\vert y_1-y_0
\vert$, is the absolute value of the difference of the extra
coordinate.

Denote with $\pi: P \to M$ the projection to $M$, with $\pi_{H}: P
\to H$ the projection to $H$, and with $y\in H$, the generic point
of $H$. The direct product structure defines at each point of $p \in
P$, a natural splitting of $TP_p$ into  horizontal and a vertical
parts, namely a generalized connection. Given a vector $V \in TP_p$,
$V$ is horizontal if $\pi_{H*}V=0$, while it is vertical if
$\pi_{*}V=0$. The horizontal lift of $v\in TM_x$, to $p \in P$,
$\pi(p)=x$, is defined as usual as the only horizontal vector $V$
which projects to $v$.

%The 1-form field $\dd y$ is a connection \cite{kobayashi63} for the
%$(\mathbb{R},+)$-principal bundle $P$.

The time orientation of $(P, \tilde{g})$ is obtained from the global
timelike vector field obtained by taking the horizontal lift of the
global, future directed, timelike  vector field for $(M,g)$.

It is easy to check that since $h$ is positive definite, the
projection of a timelike (resp. causal) vector on $TP_p$, $p \in P$,
is a timelike (resp. causal) vector on $TM_x$, $x =\pi(p)$. Note,
however, that the projection of a lightlike vector $V$ is timelike
unless $h(\pi_{H*}V,\pi_{H*}V)=0 \Rightarrow  \pi_{H*}V=0$, i.e.
unless it is the horizontal lift of its projection, in which case
the projection is lightlike too. As a consequence the projection of
a causal (resp. timelike) curve is a causal (resp. timelike) curve,
thus $p_0 \le p_1$ (resp. $p_0 \ll p_1$) implies $\pi(p_0) \le
\pi(p_1)$ (resp. $\pi(p_0) \ll \pi(p_1)$). Finally, since the
horizontal lift of a causal (resp. timelike) curve on $M$ is a
causal (resp. timelike) curve on $P$, it is
$\pi(J^{+}(p))=J^{+}(\pi(p))$ (resp. $\pi(I^{+}(p))=I^{+}(\pi(p))$).

For applications, the warped product $(P,\hat{g})$, where $\hat{g}=
g+f^2(x) h $, and $f: M \to \mathbb{R}^{+}$, is far more interesting
than the direct product, as many physical metrics are in fact
obtained by the repeated application of warped products. Sometimes
the notation $P= M\times_{f} H $ is used although the manifold $P$
does not depend on $f$.

\begin{remark}
The spacetime $P=M\times_{f} H$ with $M=(a,b)$, $-\infty\le a < b
\le +\infty$,  and $g=-\dd t^2$, also known as generalized FRW
metric, will not be considered in this work. The reason is that the
causality of these spacetimes is trivial if $(H,h)$ is a complete
Riemannian manifold. Indeed, it has been shown that $(P,g+f^2h)$ is
globally hyperbolic iff $(H,h)$ is complete \cite[Th. 3.66]{beem96}.
\end{remark}

The next theorem proves that it is not restrictive to study the
relation between the causality properties for $(M,g)$ and the direct
product $(P,\tilde{g})$ as the results hold also for the warped
product whatever the positive function $f$.

\begin{theorem} \label{mjh}
Let $(M,g)$ be a generic spacetime, $(H,h)$ a Riemannian manifold,
$\tilde{g}$ the direct product metric on $P=M\times H$, and
$\hat{g}$ a warped product metric for a positive function $f$. Let
$\mathcal{P}$ and $\mathcal{P}'$ be two given conformal invariant
properties. Consider the following logical statements
\begin{itemize}
\item[(a)]  $(M,g)$
satisfies $\mathcal{P}$ $\Rightarrow $ $ (P,\tilde{g})$ satisfies
$\mathcal{P}'$,
\item[(b)]   $(M,g)$ satisfies $\mathcal{P} \Rightarrow
(P,\hat{g})$ satisfies $\mathcal{P}'$,
\item[(c)]  $(P,\tilde{g})$ satisfies
$\mathcal{P}'$ $\Rightarrow $ $(M,g)$ satisfies $\mathcal{P}$,
\item[(d)]   $(P,\hat{g})$ satisfies
$\mathcal{P}'$  $\Rightarrow $ $(M,g)$ satisfies $\mathcal{P}$,
\end{itemize}
 Then (a) $\Leftrightarrow$ (b) and (c) $\Leftrightarrow$ (d).
\end{theorem}

\begin{proof}
(a) $\Rightarrow$ (b). Let $(M,g)$ satisfy $\mathcal{P}$, then
because $\mathcal{P}$ is a conformal invariant property,
$(M,g/f^{2})$ satisfies $\mathcal{P}$, and because of the assumed
implication, $(P,f^{-2} g+  h)$ satisfies $\mathcal{P}'$, and again
because of conformal invariance, $(P,g+ f^2 h)$ satisfies
$\mathcal{P}'$.

(b) $\Rightarrow$ (a).  Let $(M,g)$ satisfy $\mathcal{P}$, then
because $\mathcal{P}$ is a conformal invariant property, $(M, f^2
g)$ satisfies $\mathcal{P}$, and because of the assumed implication,
$(P,f^2 g+ f^2  h)$ satisfies $\mathcal{P}'$, and again because of
conformal invariance, $(P,g+ h)$ satisfies $\mathcal{P}'$.

(c) $\Rightarrow$ (d). Let $(P,\hat{g})$ satisfy $\mathcal{P}'$,
then because $\mathcal{P}'$ is a conformal invariant property,
$(P,f^{-2} g+ h )$ satisfies $\mathcal{P}'$, and because of the
assumed implication, $(M,f^{-2}g)$ satisfies $\mathcal{P}$, and
again because of conformal invariance, $(M,g)$ satisfies
$\mathcal{P}$.

(d) $\Rightarrow$ (c). Let $(P,\tilde{g})$ satisfy $\mathcal{P}'$,
then because $\mathcal{P}'$ is a conformal invariant property,
$(P,f^2 g+f^2  h)$ satisfies $\mathcal{P}'$, and because of the
assumed implication, $(M,f^2 g)$ satisfies $\mathcal{P}$, and again
because of conformal invariance, $(M,g)$ satisfies $\mathcal{P}$.

\end{proof}

Thus, assume to know, for instance, that if $(M,g)$ is
distinguishing then $(P,\tilde{g})$ is distinguishing, then the
previous theorem implies that $ M \times_{f} H$ is distinguishing
too, whatever the positive function $f$.

In what follows I shall clarify whether a result holds in the warped
product case $ M \times_{f} H$ or only in the direct product case $
M \times_{} H$. Nevertheless, the proofs will be given only for the
direct product case, the generalization to the warped product being
trivial because of the previous theorem. The theorems which will not
be generalizable to the warped product case are those which depend
on non-conformal properties of the base spacetime $(M,g)$. Usually,
these non-conformal invariant conditions are completeness
conditions, geodesic connectedness conditions or continuity
requirements on the Lorentzian distance function $d$.
The properties
of causal continuity and causal simplicity will present such a
difficulty.

From now on $(H,h)$ is a {\em complete} Riemannian manifold. The
next lemma clarifies the relation between the additional dimensions
and the Lorentzian distance on $M \times M$. In the convention of
this work the inclusion $\subset$ is a reflexive relation $U\subset
U$.
\begin{lemma} \label{uty}
Consider the direct product spacetime $(P,\tilde{g})$, $P=M\times
H$. Let $p_0=(x_0,y_0) \in P$, for every $x_1 \in M$, it is
\begin{equation} \label{iyr}
I^{+}(p_0) \cap \pi^{-1}(x_1)=\{(x_1,y): \rho(y_0,y) < d(x_0,x_1)\}
,
\end{equation}
moreover, $J^{+}(p_0) \cap \pi^{-1}(x_1) \ne \emptyset$ only if $x_1
\in J^{+}(x_0)$ and
\begin{align}
 J^{+}(p_0) \cap \pi^{-1}(x_1)&\subset  \{(x_1,y):
\rho(y_0,y) \le d(x_0,x_1)\}. \label{iyr2}
\end{align}
If $d(x_0,x_1)=+\infty$, then $E^{+}(p_0) \cap
\pi^{-1}(x_1)=\emptyset$, otherwise $d(x_0,x_1)<+\infty$ and
\begin{align}
  E^{+}(p_0) \cap \pi^{-1}(x_1)&\subset  \{(x_1,y):
\rho(y_0,y) = d(x_0,x_1)\}. \label{iyr3}
\end{align}
Analogous past versions of these statements also hold.
\end{lemma}

\begin{proof}
If the set $I^{+}(p_0) \cap \pi^{-1}(x_1)$ is not empty there is a
timelike curve connecting $p_0$ with $x_1$'s fiber. Its projection
is a timelike curve that connects $x_0$ to $x_1$ thus, $x_1 \in
I^{+}(x_0)$ or the set $I^{+}(p_0) \cap \pi^{-1}(x_1)$ is empty. The
right-hand side of Eq. (\ref{iyr}) gives an empty set if $x_1 \notin
I^{+}(x_0)$ thus there remains to consider the case $x_1 \in
I^{+}(x_0)$. Let $\sigma(\lambda)$, $\lambda \in [0,1]$ be any
$(C^1)$ timelike curve from $x_0$ to $x_1$, let $y: [0,+\infty) \to
H$, be a geodesic parametrized with respect to length starting from
$y_0$, and consider for any given $\alpha \in [-1,1]$, the curve on
$P$
\[
\gamma(\lambda)=(\sigma(\lambda), y(\alpha \int_{\sigma
x_0}^{\sigma(\lambda)} \dd s) ).
\]
It can be easily checked to be timelike for $\alpha =(-1,1)$ and
lightlike for $\vert \alpha\vert=1$. Its second endpoint is
$(x_1,y(\alpha l(\sigma)))$. Thus every event $(x_1,y_1)$, $
\rho(y_0,y_1) < d(x_0,x_1) $, can be reached by a timelike curve
from $p_0$, simply choose $\sigma$ such that $l(\sigma) >
\rho(y_0,y_1)$, the constant $\alpha$ so that $ \alpha
l(\sigma)=\rho(y_0,y_1)$, and the geodesic $y$ to be a minimizing
geodesic connecting $y_0$ to $y_1$. Finally, if $\gamma(\lambda)$ is
a timelike (resp. causal) curve from $p_0$ to $x_1$'s fiber and
$\sigma$ is its projection, the timelike (resp. causal) condition
reads $ \sqrt{ h(\frac{\dd y}{\dd \lambda}, \frac{\dd y}{\dd
\lambda})}< \frac{\dd s}{\dd \lambda}$ (resp.  $\sqrt{ h(\frac{\dd
y}{\dd \lambda}, \frac{\dd y}{\dd \lambda})} \le \frac{\dd s}{\dd
\lambda}$), so that if $(x_1,y_1)$ is its second endpoint,
$\rho(y_0,y_1) \le l_H(y) < l(\sigma)\le d(x_0,x_1)$ (resp.
$\rho(y_0,y_1) \le l_H(y) \le l(\sigma)\le d(x_0,x_1)$).
\end{proof}

The next lemma summarizes some results basically due to Beem and Powell \cite{beem82}. 
I give a simple proof for the reader convenience

\begin{lemma} \label{itg}
Consider the direct product spacetime $(P,\tilde{g})$, $P=M\times
H$. Let $\gamma(\lambda)$, $\lambda \in [0,1]$, be a causal (resp.
timelike) geodesic on $(P,\tilde{g})$ connecting $p_0=(x_0,y_0)$ to
$p_1=(x_1,y_1)$, then it decomposes as
\[
\gamma(\lambda)=(\sigma(\lambda), y(\lambda)),
\]
where $\sigma(\lambda)$ is a causal (resp. timelike) geodesic on
$(M,g)$, and $y(\lambda)$ is a geodesic on $H$ (possibly degenerated
to the only point $y_0$) connecting $y_0$ to $y_1$ of length $l_H(y)
\le l(\sigma)$ (resp. $l_H(y)  < l(\sigma)$). Moreover, if $\gamma$
is a maximizing lightlike geodesic then $\sigma$ is a maximizing
geodesic and $y$ is a minimizing geodesic.
\end{lemma}

\begin{proof}
That the projection $\sigma$ is a causal (resp. timelike) geodesic
follows from the direct product structure, and the same can be said
for the geodesic nature of $y(\lambda)$. From the condition of
causality (resp. chronology) for $\gamma$, that is  $ -(\frac{\dd
s}{\dd \lambda})^{2}+h(\frac{\dd y}{\dd \lambda},\frac{\dd y}{\dd
\lambda}) \le 0$ $\Rightarrow$ $\sqrt{ h(\frac{\dd y}{\dd \lambda},
\frac{\dd y}{\dd \lambda})} \le \frac{\dd s}{\dd \lambda}$ (resp.
$\sqrt{ h(\frac{\dd y}{\dd \lambda}, \frac{\dd y}{\dd \lambda})} <
\frac{\dd s}{\dd \lambda}$), it follows $l_H(y) \le l(\sigma)$
(resp. $l_H(y)  < l(\sigma)$). The last statement follows from
$\rho(y_0,y_1) \le l_H(y) \le l(\sigma)\le d(x_0,x_1)$, and from the
inclusion (\ref{iyr3}) because as $\gamma$ is lightlike and
maximizing, $p_1 \in E^{+}(p_0)$, thus $d(x_0,x_1)=\rho(y_0,y_1)$,
and hence $l(\sigma)=d(x_0,x_1)$, $l_H(y)=\rho(y_0,y_1)$.
\end{proof}

It is worth mentioning that Beem and Powell \cite{beem82} were able to extended
some aspects of this direct product  result to the warped product case. They proved that  in this last case the
projection $y$ on $H$ of a geodesic $\gamma$ on $P$ is a pregeodesic
of $(H,h)$, and that if $\gamma$ is maximizing then $y$ is
minimizing. However, the projection  on  $M$, $\sigma$, is not a geodesic. Although interesting, we shall need the simpler direct product case  given above in
what follows. The next theorem gives a key relation between all the
distance functions involved in a direct product spacetime.

\begin{theorem} \label{cfd}
Consider the direct product spacetime $(P,\tilde{g})$, $P=M\times
H$. Let $p_0=(x_0,y_0)$, $p_1=(x_1,y_1)$, be events in
$(P,\tilde{g})$, then if $d^{(P)}$ is the Lorentzian distance on
$P\times P$,
\begin{equation} \label{ynb}
d^{(P)}(p_0,p_1)=\sqrt{d(x_0,x_1)^2-\rho(y_0,y_1)^2} ,
\end{equation}
whenever the argument of the square root is positive, otherwise
$d^{(P)}(p_0,p_1)=0$.
\end{theorem}

\begin{proof}
If $d(x_0,x_1)\le \rho(y_0,y_1)$ by  lemma \ref{uty} $(p_0,p_1)
\notin I^{+}$, thus $d^{(P)}(p_0,p_1)=0$, as claimed.

Let $ \rho(y_0,y_1)  < d(x_0,x_1)$ so that $d(x_0,x_1) >0$, and let
$\sigma(\lambda)$, $\lambda \in [0,1]$, be a timelike curve
connecting $x_0$ to $x_1$ such that $\rho(y_0,y_1)<l(\sigma)\le
d(x_0,x_1)$. Let $y$ be a minimizing geodesic connecting $y_0$ to
$y_1$, parametrized with respect to length, and starting from $y_0$.
The curve
\[
\gamma(\lambda)=(\sigma(\lambda),y(\frac{\rho(y_0,y_1)}{l(\sigma)}\int_{\sigma
x_0}^{\sigma(\lambda)} \dd s)) ,
\]
is timelike, connects $p_0$ to $p_1$ and has length
\[
\int_{\gamma} \dd s^{(P)}=\int_{\sigma}
\sqrt{1-(\frac{\rho(y_0,y_1)}{l(\sigma)})^2}\, \dd
s=\sqrt{l(\sigma)^2-\rho(y_0,y_1)^2},
\]
thus, taking the lower upper bound over the space of timelike
connecting curves on the base
\[
d^{(P)}(p_0,p_1) \ge \sup_{\sigma}
\sqrt{l(\sigma)^2-\rho(y_0,y_1)^2}=\sqrt{d(x_0,x_1)^2-\rho(y_0,y_1)^2}.
\]
But if it were
$d^{(P)}(p_0,p_1)>\sqrt{d(x_0,x_1)^2-\rho(y_0,y_1)^2}$ there would
be a timelike curve
$\tilde\gamma(\lambda)=(\tilde\sigma(\lambda),y(\lambda))$ on $P$ of
length greater  than $\sqrt{d(x_0,x_1)^2-\rho(y_0,y_1)^2}$
\[
\sqrt{d(x_0,x_1)^2-\rho(y_0,y_1)^2}<\int_{\tilde\gamma} \dd
s^{(P)}=\int_{\tilde\sigma}  \sqrt{1-h(\frac{\dd y}{\dd s},
\frac{\dd y}{\dd s})} \, \dd s = \int_{\tilde\sigma} \sqrt{1-v^2(s)}
\, \dd s,
\]
where $v(s)= \sqrt{h(\frac{\dd y}{\dd s}, \frac{\dd y}{\dd s})}$ and
$\int_{\tilde\sigma} v(s) \dd s=l_H(y)$. It is well known from
special relativity that the right-hand side is maximized  if $v(s)$
is a constant necessarily equal to $l_H(y)/l(\tilde\sigma)$, thus
\[
\sqrt{d(x_0,x_1)^2-\rho(y_0,y_1)^2}<
\sqrt{l(\tilde\sigma)^2-l_H(y)^2} \le
\sqrt{l(\tilde\sigma)^2-\rho(y_0,y_1)^2},
\]
that is $d(x_0,x_1) <l(\tilde\sigma)$, a contradiction.

\end{proof}

\begin{lemma} \label{uty2}
Consider the direct product spacetime $(P,\tilde{g})$, $P=M\times
H$. Let $p_0=(x_0,y_0) \in P$, for every $x_1 \in M$,
$\bar{J}^{+}(p_0) \cap \pi^{-1}(x_1) \ne \emptyset$ only if  $x_1
\in \bar{J}^{+}(x_0)$, in which case
\begin{equation}
\bar{J}^{+}(p_0) \cap \pi^{-1}(x_1)=\{(x_1,y): \rho(y_0,y) \le
S^{+}(x_0,x_1)\} ,\label{iyrt}
\end{equation}
where
\begin{equation}
S^{+}(x_0,x_1)=\inf_{U\ni x_1} \sup_{x \in U} d(x_0,x)\ge
d(x_0,x_1),\label{iyrt2}
\end{equation}
($U$ open set) and $S^{+}(x_0,x_1)-d(x_0,x_1)$ is the discontinuity
of the restricted distance function $d(x_0,\cdot): M \to
[0,+\infty]$, at $x_1$.
\end{lemma}

\begin{proof}
Let $q=(x_1,y_1) \in \bar{J}^{+}(p_0) \cap \pi^{-1}(x_1) $ then
there is a sequence $q_i=(x_i,y_i) \to (x_1,y_1)$, $q_i \in
J^{+}(p_0)$, thus $x_i \in J^{+}(x_0)$ and $x_1 \in
\bar{J}^{+}(x_0)$ as claimed.

The inequality $S^{+}(x_0,x_1) \ge d(x_0,x_1)$ follows from the fact
that for every open set $U\ni x_1$, $\sup_{x \in U} d(x_0,x) \ge
d(x_0,x_1)$ because $x_1$ belongs to $U$. From the definition of
$S^{+}(x_0,x_1)$ it also follows that for every sequence $x_i \to
x_1$ then $\limsup_{i \to +\infty}d(x_0,x_i)\le S^{+}(x_0,x_1)$.
Coming back to the sequence $q_i$, since, because of lemma
\ref{uty}, $\rho(y_0,y_i) \le d(x_0,x_i)$, we have
\[ \rho(y_0,y_1) =
\limsup_{i \to +\infty} \rho(y_0,y_i) \le \limsup_{i \to
+\infty}d(x_0,x_i) \le S^{+}(x_0,x_1),
\]
and
thus $q$ stays in the set given by the right-hand side of Eq.
(\ref{iyrt}).

In order to prove the other inclusion, let $x_1 \in
\bar{J}^{+}(x_0)$ and consider the two cases $S^{+}(x_0,x_1)=0$ and
$S^{+}(x_0,x_1)>0$. In the former case, it is clear that the event
$(x_1,y_0)$ belongs to $\bar{J}^{+}(p_0) \cap \pi^{-1}(x_1)$, indeed
if $x_i \to x_1$,  $x_0<x_i$, and $\sigma_i$ is a causal curve
connecting $x_0$ to $x_i$, then its horizontal lift is a causal
curve which connects $p_0$ to $q_i=(x_i, y_0) \to (x_1,y_0)$.

In the latter case let $q=(x_1,y_1)$ with $\rho(y_0,y_1) \le
S^{+}(x_0,x_1)$. From the definition of $S^{+}(x_0,x_1)$ it is not
difficult to show that there is always a sequence $x_i \to x_1$ such
that $\lim_{i \to +\infty} d(x_0,x_i)=S^{+}(x_0,x_1)$. Clearly we
can assume $x_i \in I^{+}(x_0)$ since $S^{+}(x_0,x_1)>0$. Let
$\sigma_i(s)$ be a timelike curve connecting $x_0=\sigma_i(0)$ to
$x_i$, parametrized with proper time, and of length
$l(\sigma_i)>d(x_0,x_i)-\frac{1}{i}$ if $d(x_0,x_i)<+\infty$, or
$l(\sigma_i)>i$ if  $d(x_0,x_i)=+\infty$. With this choice $\lim_{i
\to +\infty} l(\sigma_i)=\lim_{i \to
+\infty}d(x_0,x_i)=S^{+}(x_0,x_1)$. Let $y$ be a minimizing geodesic
connecting $y_0$ to $y_1$, parametrized with respect to length and
starting at $y_0$.

If $\rho(y_0,y_1) < S^{+}(x_0,x_1)$ the  curve
\[
\gamma_i=(\sigma_i(s), y(\frac{\rho(y_0,y_1)}{l(\sigma_i)} \,s)),
\]
is causal for sufficiently large $i$ and connects $p_0$ to
$(x_i,y_1)$ whose limit is $(x_1,y_1)$, thus $(x_1,y_1) \in
\bar{J}^{+}(p_0)$.

If $\rho(y_0,y_1) =S^{+}(x_0,x_1)$ the  curve
\[
\gamma_i=(\sigma_i(s), y( s)),
\]
is lightlike and hence causal  and connects $p_0$ to $(x_i,y(
l(\sigma_i)))$ whose limit is
$(x_1,y(S^{+}(x_0,x_1)))=(x_1,y(\rho(y_0,y_1)))=(x_1,y_1)$, thus
$(x_1,y_1) \in \bar{J}^{+}(p_0)$.
\end{proof}

\begin{remark}
There is an analogous past version of lemma \ref{uty2}. Let
$p_1=(x_1,y_1) \in P$, for every $x_0 \in M$, $\bar{J}^{-}(p_1) \cap
\pi^{-1}(x_0) \ne \emptyset$ only if  $x_0 \in \bar{J}^{-}(x_1)$, in
which case
\begin{equation}
\bar{J}^{-}(p_1) \cap \pi^{-1}(x_0)=\{(x_0,y): \rho(y,y_1) \le
S^{-}(x_0,x_1)\} ,
\end{equation}
Here $S^{-}(x_0,x_1)=\inf_{U\ni x_0} \sup_{x \in U} d(x,x_1)\ge
d(x_0,x_1)$, and $S^{-}(x_0,x_1)-d(x_0,x_1)$ is the discontinuity of
the restricted distance function $d(\cdot,x_1): M \to [0,+\infty]$,
at $x_0$.

\end{remark}

Lemma \ref{uty2} will be particularly important in connection with
causal continuity. The next lemma will be useful in connection with
causal simplicity.

\begin{lemma} \label{uty3}
Consider the direct product spacetime $(P,\tilde{g})$, $P=M\times
H$. For every pair $x_0,x_1 \in M$, $\bar{J}^{+} \cap[
\pi^{-1}(x_0)\times \pi^{-1}(x_1)] \ne \emptyset$ only if $(x_0,x_1)
 \in \bar{J}^{+}$, in which case
\begin{align}
&\bar{J}^{+} \cap [ \pi^{-1}(x_0)\times \pi^{-1}(x_1)] \nonumber \\
&=\{(p_0,p_1): \ p_0=(x_0,y_0), p_1=(x_1,y_1) \textrm{ and
}\rho(y_0,y_1) \le D(x_0,x_1)\} ,
\end{align}
where
\begin{equation}
D(x_0,x_1)=\inf_{V\ni (x_0,x_1)} \sup_{(x,z) \in V} \! \! d(x,z)\,
\ge d(x_0,x_1),
\end{equation}
($V\subset M\times M$ open set) and $D(x_0,x_1)-d(x_0,x_1)$ is the
discontinuity of the distance function $d: M\times M \to
[0,+\infty]$, at $(x_0,x_1)$.
\end{lemma}

\begin{proof}
Let $(p_0,p_1)  \in \bar{J}^{+} \cap[ \pi^{-1}(x_0)\times
\pi^{-1}(x_1)] $, $p_0=(x_0,y_0)$, $p_1=(x_1,y_1)$, then there is a
converging sequence $(p_{0i},p_{1i}) \to (p_0,p_1)$ with
$(p_{0i},p_{1i}) \in J^{+}$. Denoting $p_{0i}=(x_{0i},y_{0i})$ and
$p_{1i}=(x_{1i},y_{1i})$, it follows $(x_{0i},x_{1i}) \in J^{+}$ and
since  $(x_{0i},x_{1i}) \to (x_0,x_1)$, $(x_0,x_1) \in \bar{J}^{+}$
as claimed.

The inequality $D(x_0,x_1) \ge d(x_0,x_1)$ follows from the fact
that for every open set $V\ni (x_0,x_1)$, $\sup_{(x,z) \in V} d(x,z)
\ge d(x_0,x_1)$ because $(x_0,x_1)$ belongs to $V$. From the
definition of $D(x_0,x_1)$ it also follows that for every sequence
$(x_{0i},x_{1i}) \to (x_0,x_1)$ then $\limsup_{i \to
+\infty}d(x_{0i},x_{1i})\le D(x_0,x_1)$. In particular for the
sequence $(x_{0i},x_{1i})$ constructed above, since, because of
lemma \ref{uty}, $\rho(y_{0i},y_{1i}) \le d(x_{0i},x_{1i})$, we have
\[ \rho(y_0,y_1) =
\limsup_{i \to +\infty} \rho(y_{0i},y_{1i}) \le \limsup_{i \to
+\infty} d(x_{0i},x_{1i}) \le D(x_0,x_1),
\]
and thus $(p_0,p_1)$ stays in the set given by the right-hand side
of Eq. (\ref{iyrt}).

In order to prove the other inclusion, let $(x_0,x_1) \in
\bar{J}^{+}$ and consider the two cases $D(x_0,x_1)=0$ and
$D(x_0,x_1)>0$. In the former case, we have only to prove that, for
any given $y_0$, $p_0=(x_0,y_0)$, the event $p_1=(x_1,y_0)$ is such
that $(p_0,p_1) \in \bar{J}^{+}$, which is trivial because  if
$(x_{0i},x_{1i}) \to (x_0,x_1)$, $(x_{0i},x_{1i})\in J^{+}$, and
$\sigma_i$ is a causal curve connecting $x_{0i}$ to $x_{1i}$, then
its horizontal lift is a causal curve which connects
$p_{0i}=(x_{0i},y_0)$ to $p_{1i}=(x_{1i},y_0)$, and $(p_{01},p_{1i})
\to (p_0,p_1)$.

In the latter case let $p_0=(x_0,y_0)$ and $p_1=(x_1,y_1)$ with
$\rho(y_0,y_1) \le D(x_0,x_1)$. From the definition of $D(x_0,x_1)$
it is not difficult to show that there is always a sequence
$(x_{0i},x_{1i}) \to (x_0,x_1)$ such that $\lim_{i \to +\infty}
d(x_{0i},x_{1i})=D(x_0,x_1)$. Clearly we can assume $(x_{0i},x_{1i})
\in I^{+}$ since $D(x_0,x_1)>0$. Let $\sigma_i(s)$ be a timelike
curve connecting $x_{0i}=\sigma_i(0)$ to $x_{1i}$, parametrized with
proper time, and of length
$l(\sigma_i)>d(x_{0i},x_{1i})-\frac{1}{i}$ if
$d(x_{0i},x_{1i})<+\infty$, or $l(\sigma_i)>i$ if
$d(x_{0i},x_{1i})=+\infty$. With this choice $\lim_{i \to +\infty}
l(\sigma_i)=\lim_{i \to +\infty}d(x_{0i},x_{1i})=D(x_0,x_1)$. Let
$y$ be a minimizing geodesic connecting $y_0$ to $y_1$, parametrized
with respect to length and starting at $y_0$.

If $\rho(y_0,y_1) < D(x_0,x_1)$ the  curve
\[
\gamma_i=(\sigma_i(s), y(\frac{\rho(y_0,y_1)}{l(\sigma_i)}\, s)),
\]
is causal for sufficiently large $i$ and connects $p_{0i}=(x_{0i},
y_0)$ to $p_{1i}=(x_{1i},y_1)$, moreover $(p_{0i},p_{1i}) \to
(p_0,p_1)$, thus $(p_0,p_1) \in \bar{J}^{+}$.

If $\rho(y_0,y_1) =D(x_0,x_1)$ the  curve
\[
\gamma_i=(\sigma_i(s), y( s)),
\]
is lightlike and hence causal  and connects $p_{0i}=(x_{0i},y_0)$ to
$(x_{1i},y( l(\sigma_i)))$ whose limit is
$(x_1,y(D(x_0,x_1)))=(x_1,y(\rho(y_0,y_1)))=(x_1,y_1)=p_1$, while
$p_{0i} \to p_0$, thus $(p_0,p_1) \in \bar{J}^{+}$.
\end{proof}

\begin{theorem} \label{pcr}
Whatever the value of $f$, $P=M\times_f H$, the warped product
spacetime $(P,\bm{\hat{g}})$ is chronological (resp. causal,
strongly causal, stably causal, globally hyperbolic) iff
$(M,\bm{g})$ is chronological (resp. causal, strongly causal, stably
causal, globally hyperbolic).
\end{theorem}

\begin{proof}
Recall that the proof can be given in the more specialized direct
product case because the involved properties are conformal invariant
(see theorem \ref{mjh}). This result is proved for instance in
\cite[Prop. 3.61,3.62,3.64,3.68]{beem96} (for the globally
hyperbolic case see also \cite{walschap95,caponio03,minguzzi06}). I
give explicitly the proof for the causal case as we will use it. If
$(P,\bm{\tilde{g}})$ is not causal then there is a closed causal
curve whose projection is a closed causal curve for $(M,\bm{g})$.
Conversely, if $(M,\bm{g})$ admits a closed causal curve then its
horizontal lift is a closed causal curve for $(P,\bm{\tilde{g}})$.
\end{proof}

Recall that a spacetime $(M,g)$ is totally vicious if for every pair
of events $x,z \in M$, $d(x,z)=+\infty$.

\begin{theorem}
Whatever the value of $f$, $P=M\times_f H$, the warped product
spacetime $(P,\bm{\hat{g}})$ is non-totally vicious iff $(M,\bm{g})$
is non-totally vicious.
\end{theorem}

\begin{proof}
Recall that the proof can be given in the more specialized direct
product case because the involved property is conformal invariant
(see theorem \ref{mjh}). In this case the claim follows easily from
Eq. (\ref{ynb}).
\end{proof}

Let us consider the distinguishing property. We need some
preliminary results.

\begin{lemma} \label{kijd}
Let $(M,g)$ be a spacetime, if $I^{+}(x_2)\subset I^{+}(x_1)$, $x_1
\ne x_2$, then for every $z \in I^{+}(x_2)$, $d(x_2,z) \le
d(x_1,z)$. In particular, if $I^{+}(x_1)=I^{+}(x_2)$, $x_1 \ne x_2$,
then for every $z \in I^{+}(x_1)$, $d(x_1,z)=d(x_2,z)$. Analogous
past versions of these statements also hold.
\end{lemma}

\begin{proof}
 Indeed, let $\sigma_2(s)$ be
a timelike curve connecting $x_2$ to $z$ and let it be parametrized
with respect to proper time. For every $0<\epsilon < l(\sigma_2)$,
$\sigma_2(\epsilon) \in I^{+}(x_1)$ thus there is a timelike curve
$\sigma_1$ which connects first $x_1$ to $\sigma_2(\epsilon)$ and
then this event to $z$ following $\sigma_2$. Thus
$l(\sigma_1)+\epsilon \ge l(\sigma_2)$, and taking the sup over the
set of connecting timelike curves $\sigma_2$, $d(x_1,z)+\epsilon \ge
d(x_2,z)$. As $\epsilon$ is arbitrary $d(x_1,z)\ge d(x_2,z)$, and
analogously in the other direction by interchanging the roles of
$x_1$ and $x_2$.
\end{proof}

The next result has an analog in the strongly causal case \cite[Cor.
4.28]{beem96}.

\begin{theorem} \label{xdy}
If $(M,g)$ is a future (resp. past) distinguishing spacetime then
for any $x\in M$, there is an arbitrary small open neighborhood
$U_x$ (which can be chosen globally hyperbolic) such that $d(x,
\cdot): U_x \to [0,+\infty]$ is continuous and finite (resp.
$d(\cdot,x): U_x \to [0,+\infty]$ is continuous and finite).
\end{theorem}

\begin{proof}
Since $M$ is future distinguishing \cite[Lemma 3.10]{minguzzi06c}
for every open set $U\ni x$ there is a neighborhood $V \subset U$,
$V\ni x$ such that every timelike curve starting from $x$ and ending
at $y \in V$, is necessarily contained in $V$. Moreover, the same
proof \cite[Lemma 3.10]{minguzzi06c} shows that $V$ can be chosen
globally hyperbolic when regarded as a spacetime with the induced
metric. As a consequence $d(x, \cdot): V \to [0,+\infty]$ coincides
with $d\vert_{V\times V}(x,\cdot)$, where $ d\vert_{V\times V}$ is
the Lorentzian distance on the spacetime $(V,g\vert_{V})$. Since $V$
is globally hyperbolic $d\vert_{V\times V}(x,\cdot)$ is continuous
and finite and so is $d(x, \cdot): V \to [0,+\infty]$.
\end{proof}

The next example shows that the assumption of distinction is needed
in the previous lemma.

\begin{example}
Consider the spacetime $M=\mathbb{R}\times S^1\backslash \{o\}$, of
coordinates $(t,\theta)$, $\theta \in (-\pi,+\pi]$, $o=(0,0)$,  and  metric $g=-\omega^{-} \otimes \omega^{+}-\omega^{+} \otimes \omega^{-}$, where $\omega^{-}=\cos\alpha(t)\dd t - \sin \alpha(t) \dd \theta $ and $\omega^{+}=\sin\alpha(t)\dd t + \cos \alpha(t) \dd \theta$, $\alpha(t)\ge 0 $, $\alpha(t)$   even function with $ \alpha'> 0$ in $(0,1)$,  $\alpha(t)=\pi/4$ if  $\vert t\vert \ge 1$, see
figure \ref{dist}.

\begin{figure}
\centering
\includegraphics[width=2.2cm]{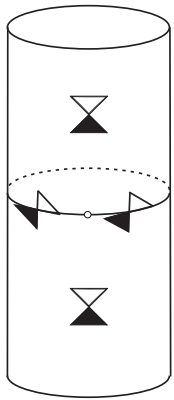}
\caption{A causal but non-distinguishing spacetime. A suitable
metric compatible with the drawn conformal structure exists which
makes the restricted Lorentzian distance  $d(x, \cdot)$ neither
finite nor continuous in neighborhoods of the events $x$ belonging
to the middle circle.} \label{dist}
\end{figure}

Let $x=(0,\phi)$, $\phi\ne 0$, be a point in the middle circle, and let $z$ be a point such that $t(z)>0$.
Consider the past inextendible timelike curve $\sigma$ of future endpoint $z$ and of equation $\dd t/\dd \theta=\tan \beta(t)$ where  $\tan\beta=\tan\alpha+t^2$. As $\alpha<\beta<\pi/2$ the curve $\sigma$ is indeed timelike.
Because of the differential equation it satisfies, the curve can not cross the middle circle, as a consequence it passes infinitely often on any neighborhood of $x$. Assume $\tan\alpha(t)=t^2$ for small $\vert t\vert$.
The curve $\sigma$ can be parametrized with $t$ and it has infinite length because the integral
\[
 \int_{\epsilon}^{t(z)} \frac{\dd \tau}{\dd t} \dd t\ge \int_{\epsilon}^{t(z)}\sqrt{2\cos^2\alpha(t) (1-\tan\alpha(t) \frac{\dd \theta}{\dd t}) \frac{\dd \theta}{\dd t}}\,\dd t \sim \frac{1}{\sqrt{2}}\int_{\epsilon}^{t(z)} \frac{1}{t} \dd t ,
\]
diverges as  $\epsilon \to 0^{+}$. As a consequence, for suffieciently large $n$, we can choose to cut the curve  near $x$ so that it takes $n$ cycles around the  cylinder to get to $z$, and then to join $x$ with a causal curve to the starting point of the found segment. The found sequence of causal curves $\gamma_n$ connecting $x$ to $z$ has increasing length which goes to infinity with $n$, thus  $d(x,z)=+\infty$, where $z$  is a generic point with $t(z)>0$.

Thus the
Lorentzian distance is neither finite nor continuous in a
neighborhood of $x$ as  $d(x,x)=0$. 
\end{example}

\begin{theorem} \label{pcr2}
Whatever the value of $f$, $P=M\times_f H$, the warped product
spacetime $(P,\bm{\hat{g}})$ is  (future, past) distinguishing iff
$(M,\bm{g})$ is (future, past) distinguishing.
\end{theorem}

\begin{proof}

Since the distinction property is conformally invariant the proof
can be restricted to the direct product spacetime case, see theorem
\ref{mjh}. Assume $(M,g)$ is not future distinguishing, then there
are $x_1$, $x_2\in M$, $x_1 \ne x_2$, such that
$I^{+}(x_1)=I^{+}(x_2)$. Defined $p_1=(x_1,y_0)$ and
$p_2=(x_2,y_0)$, by lemma \ref{kijd} and lemma \ref{uty},
$I^{+}(p_1)=I^{+}(p_2)$. Thus if $(P,{\tilde{g}})$ is future
distinguishing then $(M,g)$ is future distinguishing. Conversely, if
$(M,g)$ is future distinguishing, assume that there exist $p_1\ne
p_2$, $I^{+}(p_1)=I^{+}(p_2)$. It follows $I^{+}(x_1)=I^{+}(x_2)$,
and since $(M,g)$ is future distinguishing $x_1=x_2$, thus $p_1$ and
$p_2$ stay in the same fiber. But since $(M,g)$ is chronological
$d(x_1,x_1)=0$ and since $p_1 \ne p_2$ lemma \ref{uty2} implies that
there is a discontinuity in the distance $d(x_1, \cdot)$ at $x_1$,
in contradiction with theorem \ref{xdy}.
\end{proof}

\begin{lemma}
The direct product spacetime $(P,\tilde{g})$ is geodesically
connected if and only if $(M,g)$ is geodesically
 connected.
\end{lemma}

\begin{proof}
Trivial taking into account that any geodesic $\gamma(\lambda)$  can
be written
\[
\gamma(\lambda)=(\sigma(\lambda),\, y(\lambda)) , \] where both
$\sigma(\lambda)$ and $y(\lambda)$ are geodesics, and taking into
account that $(H,h)$ is geodesically connected.

\end{proof}

\begin{definition}
A spacetime $(M,g)$ will be said to be {\em maximizing geodesically
connected} if for every pair of events, $x_1 \in
J^{+}(x_0)\backslash\{x_0\}$, $d(x_0,x_1)<+\infty$, there is a
maximizing causal geodesic $\sigma$ connecting the two events,
$l(\sigma)=d(x_0,x_1)$.
\end{definition}

 The next result, with a different (unpublished)
proof and for the special case $(H,h)=(\mathbb{R},\dd y^2)$, has
also been obtained by M. S\'anchez.

\begin{lemma} \label{cxz}
 The direct product spacetime $(P,\tilde{g})$ is maximizing geodesically connected if and only if $(M,g)$ is maximizing geodesically
 connected.
\end{lemma}

\begin{proof}
Assume $(M,g)$ is maximizing geodesically connected. Let $p_0,p_1
\in P$, $p_0=(x_0,y_0)$, $p_1=(x_1,y_1)$, with $p_0 <p_1$. Then,
since the projection of a causal curve is a causal curve, $x_0 \le
x_1$. Let $\sigma(\lambda)$, $\lambda \in [0,1]$, be the maximizing
geodesic connecting $x_0$ to $x_1$. If it is lightlike then
$d(x_0,x_1)=0$, thus $y_1=y_0$ due to the inclusion (\ref{iyr2}), in
particular there is no timelike curve connecting $p_0$ to $p_1$ thus
$d^{(P)}(p_0,p_1)=0$. The  horizontal lift of $\sigma$ gives the
maximizing lightlike geodesic.

Otherwise $d(x_0,x_1)>0$, let $y$ be a minimizing geodesic
parametrized with respect to length starting at $y_0$ and ending at
$y_1$. The curve on $P$
\[
\gamma(\lambda)=(\sigma(\lambda),y(\frac{\rho(y_0,y_1)}{d(x_0,x_1)}
\int_{\sigma\, x_0}^{\sigma(\lambda)} \dd s))
\]
is causal and connects $p_0$ to $p_1$. Its length is $\int_{\gamma}
\dd s^{(P)}=\int_{\sigma}
\sqrt{1-(\frac{\rho(y_0,y_1)}{d(x_0,x_1)})^2}\, \dd
s=\sqrt{d(x_0,x_1)^2-\rho(y_0,y_1)^2}$ and by theorem \ref{cfd},
$\gamma$ is a maximizing geodesic.

Conversely, let $(P,\tilde{g})$ be maximizing geodesically
connected. Let $x_0 <x_1$ and take $p_0=(x_0,y_0)$, $p_1=(x_1,y_0)$.
The events $p_0$, $p_1$, are connected by the horizontal lift of any
causal curve connecting $x_0$ to $x_1$, thus $p_0 <p_1$. By lemma
\ref{itg} the maximizing geodesic connecting $p_0$ to $p_1$ reads
$\gamma(\lambda)=(\sigma(\lambda), y_0)$ so that
$l^{(P)}(\gamma)=l(\sigma)$ and by theorem \ref{cfd},
$d^{(P)}(p_0,p_1)=d(x_0,x_1)$, from which it follows
$l(\sigma)=d(x_0,x_1)$, i.e. $\sigma$ is a maximizing connecting
geodesic.
\end{proof}

\begin{lemma} \label{comp}
The direct product spacetime $(P,\tilde{g})$ is (past, future)
(timelike, causal) geodesically complete if and only if $(M,g)$ is
(resp. past, future) (resp. timelike, causal) geodesically complete.
\end{lemma}

\begin{proof}
Assume $M$ is geodesically complete. Given  a geodesic
$\gamma(\lambda)=(\sigma(\lambda),y(\lambda))$ on $P$, a complete
extension of the projection $\sigma(\lambda)$ exists ($\sigma$ has
the same causal character of $\gamma$), and the same can be said for
$y(\lambda)$ as $(H,h)$ is complete. Thus the original geodesic can
be extended to a complete geodesic using the same decomposition.
Given $\sigma$ on $M$, the converse is proved taking the projection
of the extension of its horizontal lift.
\end{proof}

\section{Causal continuity and causal simplicity}
So far there has been a complete correspondence between causal
properties of $(M,g)$ and causal properties of the warped product
spacetime  $(P,\hat{g})$. Only the levels of the causal ladder
corresponding to causal continuity and causal simplicity were not
included in the previous analysis. Indeed, as we shall see, such a simple correspondence does not hold for these properties. Interestingly, they are also the
only levels of the causal ladder which are not preserved by causal
mappings \cite{garciaparrado03,garciaparrado05,garciaparrado05b}.

Recall that a causally continuous spacetime $(M,\bm{g})$ is a distinguishing
spacetime which is moreover reflecting.

\begin{theorem} \label{causc}
The direct product spacetime $(P,{\tilde{g}})$ is causally
continuous if and only if $(M,{g})$ is causally continuous and for
every pair $x_1 \in \bar{J}^{+}(x_0)$ (or equivalently, by causal
continuity, $x_0 \in \bar{J}^{-}(x_1)$),
$S^{+}(x_0,x_1)=S^{-}(x_0,x_1)$ or
\begin{equation} \label{inc}
\textrm{diam}(H,h) \le \textrm{min}[S^{+}(x_0,x_1), S^{-}(x_0,x_1)]
< +\infty.
\end{equation}

In particular, if $(M,{g})$ is causally continuous and $d:M\times M
\to [0,+\infty]$ is continuous (at least) at the pairs of events
belonging to the set $d^{-1}(C) \subset M \times M$ where
$C=[0,\textrm{diam}\, (H,h))$, then the direct product spacetime
$(P,{\tilde{g}})$ is causally continuous.
\end{theorem}

\begin{proof}
Assume $(P,\tilde{g})$ is causally continuous. By theorem \ref{pcr2}
$(M,g)$ is distinguishing, thus we have only to prove that it is
past reflective, the future case being similar. If $(M,g)$ is not
past reflecting there are events $x,z,w,$ such that $I^{+}(x)
\supset I^{+}(z)$ but $w \in I^{-}(x)$ while $w \notin I^{-}(z)$. In
particular, $d(w,z)=0$. Define $p=(x,y_0)$, $q=(z,y_0)$, $r=(w,y_0)$
for an arbitrary $y_0 \in H$. Let $q' \in I^+(q)$ then $z'=\pi(q')
\in I^{+}(z)$ thus $z' \in I^{+}(x)$. By lemma \ref{kijd},
$\rho(\pi_H(q'),y_0)<d(z,z') \le d(x,z')$ and because of lemma
\ref{uty}, $q' \in I^{+}(p)$, hence $I^{+}(p) \supset I^{+}(q)$. As
the horizontal lift of a timelike curve is a timelike curve $r \in
I^{-}(p)$, but  $r \notin I^{-}(q)$ otherwise $w \in I^{-}(z)$ hence
$(P,{\tilde{g}})$ is not past reflective, a contradiction.

Let $x_1 \in \bar{J}^{+}(x_0)$. Note that if the equality
$S^{+}(x_0,x_1)=S^{-}(x_0,x_1)$ does not hold then necessarily
$\textrm{min}[S^{+}(x_0,x_1), S^{-}(x_0,x_1)] < +\infty$. Let us
show that if $\textrm{diam}(H,h) > \textrm{min}[S^{+}(x_0,x_1),
S^{-}(x_0,x_1)]$ then $S^{+}(x_0,x_1)=S^{-}(x_0,x_1)$.

Indeed, if the equality were not satisfied, for instance
$S^{+}(x_0,x_1)>S^{-}(x_0,x_1)$, taken $y_0 \ne y_1$ such that
$S^{-}(x_0,x_1)<\rho(y_0,y_1)< \textrm{min}[\textrm{diam}\,
H,S^{+}(x_0,x_1)]$ the event $p_1=(x_1, y_1)$ would belong to
$\bar{J}^{+}(p_0)$ with $p_0=(x_0,y_0)$ by lemma \ref{uty2} but $p_0
\notin \bar{J}^{-}(p_1)$ by the same lemma, thus $(P,\tilde{g})$
would not be causally continuous (recall the definition of
reflectivity \cite[Lemma 3.42]{minguzzi06c}).

For the converse since $(M,g)$ is distinguishing, by theorem
\ref{pcr2}, $(P,{\tilde{g}})$ is distinguishing, and we have only to
show that it is reflective. Let  $p_1=(x_1,y_1) \in
\bar{J}^{+}(p_0)$, $p_0=(x_0,y_0)$, we have to show that $p_0\in
\bar{J}^{-}(p_1)$ (this fact would prove past reflectivity, the
proof for future reflectivity being analogous). But $p_1 \in
\bar{J}^{+}(p_0)$ implies $x_1 \in \bar{J}^{+}(x_0)$, and by lemma
\ref{uty2} $\rho(y_0,y_1) \le S^{+}(x_0,x_1)$ and by causal
continuity on the base $x_0 \in \bar{J}^{-}(x_1)$. There are two
cases. If the equality $S^{+}(x_0,x_1)=S^{-}(x_0,x_1)$ holds then
$\rho(y_0,y_1)\le S^{-}(x_0,x_1)$, and by lemma \ref{uty2}, $p_0\in
\bar{J}^{-}(p_1)$. Otherwise, $\rho(y_0,y_1)\le \textrm{diam}(H,h)
\le S^{-}(x_0,x_1)$ and by lemma \ref{uty2}, $p_0\in
\bar{J}^{-}(p_1)$. In conclusion $(P,\tilde{g})$ is past reflecting.

The last statements follows  because given $x_1 \in
\bar{J}^{+}(x_0)$  if  $d(x_0,x_1) <\textrm{diam}\, (H,h)$  then $d$
is continuous and hence $S^{+}(x_0,x_1)=d(x_0,x_1)=S^{-}(x_0,x_1)$.
Thus  $S^{+}(x_0,x_1)\ne S^{-}(x_0,x_1)$ only if
 $d(x_0,x_1) \ge \textrm{diam}\, (H,h)$, in which case
(recall $S^{\pm}(x_0,x_1)\ge d(x_0,x_1)$) inequality (\ref{inc})
holds.  In both cases the condition needed for the causal continuity
of the direct product is satisfied.

%
%Assume $(M,g)$ is causally continuous. By lemma \ref{pcr2}
%$(P,\tilde{g})$ is distinguishing, thus we have only to prove that
%it is past reflective, the future case being similar. If
%$(P,\tilde{g})$ is not past reflective there are events $p,q,r$,
%such that $I^{+}(p) \supset I^{+}(q)$ but $r \in I^{-}(p)$ while $r
%\notin I^{-}(q)$. As a consequence $q \in \dot{I}^{+}(p)$. Define
%$x=\pi(p)$, $z=\pi(q)$, $y=\pi(r)$. Since $\pi(I^{+}(p))=I^{+}(x)$,
%$\pi(I^{+}(q))=I^{+}(z)$, it is $I^{+}(x) \supset I^{+}(z)$, $y \in
%I^{-}(x)$.  Since $(M,g)$ is past reflective $I^{-}(z) \supset
%I^{-}(x)$, and by lemma \ref{kijd}, $d(y,x) \le d(y,z)$.  As $q \in
%\dot{I}^{+}(p)$, for a suitable choice of sign $q=(z, y_0\pm
%d(x,z))$ where $p=(x,y_0)$. By lemma \ref{uty},
%$d(y,z)<d(y,x)+d(x,z)$, otherwise $r \in I^{-}(q)$, which also
%implies that $d(y,z)<+\infty$. Finally, $0 < d(x,z)$ thus $(x,z) \in
%I^{+}$, and the reversed triangle inequality can now be applied to
%$y,x,z$, to obtain $d(y,x)+d(x,z)\le d(y,z)$. The contradiction
%proves that $(P,\tilde{g})$ is causally continuous.

\end{proof}

\begin{corollary}
Whatever the value of $f$, $P=M\times_f H$, if the warped product
spacetime $(P,\bm{\hat{g}})$ is causally continuous then
$(M,\bm{g})$ is causally continuous.
\end{corollary}

Apart  from the one already considered, a way to prove that in the case $\textrm{diam}\,
(H,h)=+\infty$ the causal continuity of $(M,g)$ and the continuity of $d$ implies the causal continuity of $(P,\tilde{g})$    is as follows.  In the first step the next result which 
improves a result by Beem and Ehrlich
\cite[Th. 4.24]{beem96} \cite{beem77}, as it imposes continuity of
$d$ only on the  vanishing distance set, is  obtained.

\begin{theorem} \label{cdfq}
Let $(M, \bm{g})$ be a conformal structure, and let a representative
$g$ exist such that  $d: M\times M \to [0,+\infty]$ is continuous on
the vanishing distance set $I^{+C}$, then $(M,\bm{g})$ is a
reflecting spacetime.
\end{theorem}

\begin{proof}
If $(M,g)$ were not reflecting then it would not be either past or
future reflecting. We can assume the first possibility as the other
case can be treated similarly. Thus there is a pair $(x,z)$ and an
event $y$ such that $I^{+}(x) \supset I^{+}(z)$ but $y \in I^{-}(x)$
while $y \notin I^{-}(z)$. In particular, $d(y,z)=0$. Since
$I^{+}(z)\subset I^{+}(x)$, $z \in \bar{I}^{+}(x)$. Let $z_n \to z$,
$z_{n} \in I^{+}(x)$, then
\[
d(y,z_n)\ge d(y,x)+d(x,z_{n})>d(y,x)>0,
\]
thus there is a discontinuity at $(y,z)$, where $d(y,z)=0$, a
contradiction.
\end{proof}

this result has the consequence

\begin{corollary} \label{fyu}
Let $(M,\bm{g})$ be a distinguishing spacetime and let a
representative $(M,g)$ exist such that the Lorentzian distance $d$
is continuous on the vanishing distance set $I^{+C}$, then
$(M,\bm{g})$ is causally continuous.
\end{corollary}

then, under the said assumptions, theorem \ref{pcr2} proves that
$(P,{\tilde{g}})$ is distinguishing, Eq. (\ref{ynb}) proves that
$d^{(P)}$ is continuous and from corollary \ref{fyu} the thesis
follows.

\begin{example}
In order to construct an example of causally continuous spacetime
$(M,g)$ such that $(P,\tilde{g})$ is not causally continuous one has
only take $(H,h)=(\mathbb{R},\dd y^2)$, so that $\textrm{diam}\,
(H,h)=+\infty$, and to find a causally continuous spacetime $(M,g)$
in which $S^{+}(x_0,x_1)\ne S^{-}(x_0,x_1)$. To this end let $M$ be
$\mathbb{R}^2$ without the origin, let $(t,x)$ be coordinates on
$\mathbb{R}^2$, and consider the metric $g=\Omega^2(-\dd t^2+\dd
x^2)$ where $\Omega^2=\frac{1}{t^{2}+\omega(x)}$,  $\omega(x)=x^2$
for $x>0$, and $\omega(x)=4x^2$ for $x<0$. Chosen $x_0=(-1,1)$,
$x_1=(1,-1)$,  the reader
may convince him or herself that $S^{+}(x_0,x_1)\ne S^{-}(x_0,x_1)$
(although a rigorous proof would require much more effort).
\end{example}

Recall that a spacetime $(M,g)$ is causally simple if it is causal
\cite{bernal06b} and such that for every $x\in M$, $J^{+}(x)$ and
$J^{-}(x)$, are closed (or, equivalenty, if it is causal and $J^{+}
\subset M \times M$ is closed \cite[Prop. 3.68]{minguzzi06c})

\begin{theorem} \label{csim}
Whatever the value of $f$, $P=M\times_f H$, if the warped product
spacetime $(P,\bm{\hat{g}})$ is causally simple then $(M,\bm{g})$ is
causally simple.
\end{theorem}

\begin{proof}
The proof can be given in the direct product case, $f=1$. It has
been already proved that $(M,g)$ is causal. Let $z \in \bar{J}^+(x)$
then there is a succession of points $z_{i} \to z$ such that $z_i
\in J^{+}(x)$ and let $\sigma_i(\lambda)$ be a causal curve
connecting $x$ with $z_i$ parametrized between 0 and 1. The
horizontal lifts $\sigma_i^{*}$ staring from $p=(x,y_0)$, $y_0 \in
H$, obtained by setting $\pi_{H}(\sigma_i^{*}(\lambda))=y_0$ are
causal and their endpoints $q_{i}=(z_i,y_0)$ converge to
$q=(z,y_0)$, thus since $P$ is causally simple there is a causal
curve $\gamma(\lambda)$ connecting $p$ and $q$. Since the projection
of a (timelike) causal curve is a (resp. timelike) causal curve the
causal curve $\sigma(\lambda)=\pi \circ \gamma$ connects $x$ and
$z$. Thus we have proved $\bar{J}^{+}(x)=J^{+}(x)$ for arbitrary $x$
and the past case is proved similarly.
\end{proof}

Recall \cite[Def. 11.1]{beem96}
\begin{definition}
The {\em timelike diameter} is, $\textrm{diam}
(M,g)=\sup\{d(x,z):x,z \in M\}$
\end{definition}

\begin{lemma} \label{sim}
If the direct product spacetime $(P,\bm{\tilde{g}})$ is causally
simple then any pair of causally related events on $M$,  $(x_0,x_1)
\in J^{+}$, such that $d(x_0,x_1) <+\infty$ and $d(x_0,x_1) \le
\textrm{diam}(H,h)$ is connected by a maximizing causal geodesic.

In particular, if $\textrm{diam}(M,g) \le \textrm{diam}(H,h)$ and
$(P,\bm{\tilde{g}})$ is causally simple then $(M,g)$ is maximizing
geodesically connected.
\end{lemma}

\begin{proof}
Let $(x_0,x_1) \in J^{+}$ such that $d(x_0,x_1) <+\infty$ and
$d(x_0,x_1) \le\textrm{diam}(H,h)$. We can find two points $y_0$ and
$y_1$ on $H$ such that $\rho(y_0,y_1)=d(x_0,x_1)$ (because of the
Hopf-Rinow theorem the distance on $H$ attains the supremum). By
lemmas \ref{uty} and \ref{uty2}, $p_1 =(x_1,y_1)$ belongs to
$E^{+}(p_0)$ where $p_0=(x_0,y_0)$, thus there is a lightlike
geodesic $\gamma(\lambda)$, $\lambda \in [0,1]$, connecting $p_0$ to
$p_1$. By lemma \ref{itg} is has the form
\[
\gamma(\lambda)=(\sigma(\lambda), y(\lambda)),
\]
where $\rho(y_0,y_1)\le l_H(y) \le l(\sigma) \le d(x_0,x_1)$  thus
in particular $l(\sigma)=d(x_0,x_1)$, that is, $\sigma$ is a
maximizing geodesic.

%The last statement follows from the fact that if
%$d(x,z)=\textrm{diam}(M,g) $ then  $d(x,z)=+\infty$ (see \cite[Prop.
%11.3]{beem96}), so that either $d(x,z)<\textrm{diam}(M,g) \le
%\textrm{diam}(H,h)$, and the previous argument holds, or
%$d(x,z)=+\infty$ and no condition is required from the property of
%maximizing geodesic connectedness.
\end{proof}

\begin{figure}
\centering  \psfrag{A}{$x$} \psfrag{B}{$t$} \psfrag{M}{$M$}
\psfrag{C}{$x_0$} \psfrag{D}{$x_1$} \psfrag{R}{Remove}
\includegraphics[width=5cm]{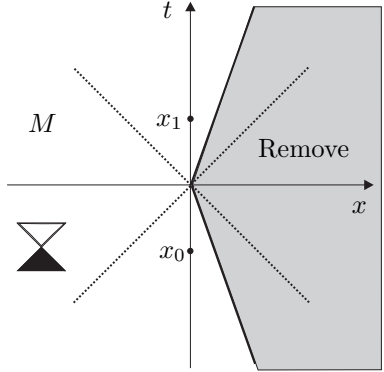}
\caption{A causally simple spacetime is not necessarily maximizing
geodesically connected. If $(M,g)$ is causally simple $(M\times
\mathbb{R},g+\dd y^2)$ is not necessarily causally simple.}
\label{fig1}
\end{figure}

In particular the spacetime $(P,\tilde{g})$ with
$(H,h)=(\mathbb{R},\dd y^2)$, constructed above the manifold $M$ of
figure \ref{fig1} is not causally simple. Indeed, given a point
$p=(x_0,0)$ in the fiber of $x_0$ the point $q=(x_1, d(x_0,x_1))$
belongs to the boundary of the causal future of $p$ but is not
causally related to it. In other words

\begin{remark}
If $(M,g)$ is causally simple $(P,\tilde{g})$ is not necessarily
causally  simple.
\end{remark}

\begin{theorem} \label{pkh}
 Let $(M,\bm{g})$ be causally simple then for any
representative $(M,g)$ the Lorentzian distance $d$ is continuous on
the vanishing distance set $I^{+C}$.
\end{theorem}

\begin{proof}
Let $d(x,z)=0$ and let $(x,z)$ be a discontinuity point for $d$,
then there is a $\epsilon>0$ and a sequence $(x_n,z_n) \to (x,z)$,
such that $d(x_n,z_n)>\epsilon>0$. In particular $(x_n,z_n) \in
I^{+}$ and $(x,z) \in \bar{I}^{+}\backslash
I^{+}=\dot{I}^{+}=E^{+}$, by causal simplicity \cite[Lemma
3.67]{minguzzi06c}. Let $\sigma_n$ be causal curves connecting $x_n$
to $z_n$ and such that $\limsup_{n \to +\infty} l(\sigma_n)
\ge\epsilon$ (for instance let $l(\sigma_n)> d(x_n,z_n)-\frac{1}{n}$
if $d(x_n,z_n)<+\infty$ and $l(\sigma_n) >n$ if
$d(x_n,z_n)=+\infty$). By \cite[Prop. 3.31]{beem96} there is a
causal curve $\gamma$ passing through $x$ and a distinguishing
subsequence $\sigma_j$ which converges to it. But by construction
any event $y\ne x,z$ of $\gamma$ is the limit of events $y_j \in
\sigma_j$, $ (x_j, y_j) \in J^{+}$, hence $(x,y) \in
\bar{J}^{+}=J^{+}$ and analogously $(y,z) \in J^{+}$, thus $\gamma$
must be a lightlike geodesic connecting $x$ to $z$, otherwise $(x,z)
\in I^{+}$. Finally,
\[
d(x,z) \ge l(\gamma) \ge \limsup_{j \to +\infty} l(\sigma_j)\ge
\epsilon>0.
\]
The contradiction concludes the proof.

\end{proof}

Thus, if $(M,g)$ is causally simple,  any discontinuity point
$(x,z)$ for the Lorentzian distance satisfies $0<d(x,z)<+\infty$.

Due to lemma \ref{sim} it is natural to look for conditions on
$(M,g)$ that guarantee the causal simplicity of $(P,\tilde{g})$. A
useful observation is

\begin{lemma} \label{bhy}
Let $(M,g)$ be a  spacetime, $P=M\times H$ the direct product
spacetime, $p_0,p_1 \in P$ and set $x_0=\pi(p_0)$, $x_1=\pi(p_1)$.
If $p_1 \in \bar{J}^{+}(p_0)\backslash {J}^{+}(p_0) (\subset
\dot{I}^{+}(p_0) =\dot{J}^{+}(p_0))$ then $d(x_0,x_1) \le
\rho(y_0,y_1)$ and one of the following three possibilities holds
\begin{itemize}
\item[(a)] $x_1 \in \bar{J}^{+}(x_0)\backslash{J}^{+}(x_0)$, \\
\item[(b)] $x_1\in J^{+}(x_0) \backslash {I}^{+}(x_0)$, and $d(x_0,\cdot):
M
\to [0, +\infty] $ is discontinuous at $x=x_1$,
\item[(c)] $x_1 \in
I^{+}(x_0)$.
\end{itemize}
Moreover, if $(M,\bm{g})$ is causally simple, only the  possibility
(c) holds.
\end{lemma}

\begin{proof}
Let $p_0=(x_0,y_0)$, $p_1=(x_1,y_1)$. By lemma \ref{uty2}  $x_1 \in
\bar{J}^{+}(x_0)$. It can not be $d(x_0,x_1)> \rho(y_0,y_1)$
otherwise by lemma \ref{uty}, $p_1 \in I^{+}(p_0)$, a contradiction.
Assume that the case (a)  and (c) do not hold and, thus, consider
the remaining case $x_1 \in J^{+}(x_0)\backslash {I}^{+}(x_0)$.

If $x_1 \in J^{+}(x_0)\backslash {I}^{+}(x_0) \, (\subset
\dot{J}^{+}(x_0))$, it is $d(x_0,x_1)=0$. If  $d(x_0,\cdot)$ were
continuous at $x_1$ then $S^{+}(x_0,x_1)=d(x_0,x_1)=0$, and by lemma
\ref{uty2}, $\rho(y_0,y_1)=0$ thus $y_1=y_0$. However, since $x_1
\in E^{+}(x_0)$ there is a lightlike geodesic connecting $x_0$ to
$x_1$ and its horizontal lift gives a causal curve connecting $p_0$
to $p_1$. The contradiction proves that $d(x_0,\cdot)$ is
discontinuous.

%There is a lightlike geodesics $\sigma(\lambda)$ without conjugate
%points before $x_1$ that connects $x_0$ to $x_1$ and that maximizes
%the length. The possibility  $y_1=y_0$ is ruled out since otherwise
%the horizontal lift of $\sigma$ would connect $p_0$ and $p_1$
%whereas by assumption $p_1 \notin J^{+}(p_0)$. The timelike curves
%$\gamma_i=(\sigma_i,y_i)$ that connect $p_0$ with $q_i$ project into
%timelike curves $\sigma_i$ which satisfy $l(\sigma_i)
%> l_H(y_i)\ge \rho(y_0,w_i)$. Thus since $w_i \to y_1 \ne y_0$
%there is  a natural number $N$ such that for $i>N$, $d(x_0,z_i) \ge
%l(\sigma_i)\ge \rho(y_0,y_1)/2>0$. Thus $d(x_0,\cdot)$ is
%discontinuous at $x_1$.

If $M$ is causally simple this case is ruled out due to lemma
\ref{pkh} (case (a) is ruled out because of the definition of causal
simplicity).

%
%We have only to prove that $d(x_0,x_1)$ is finite. Assume  that
%$d(x_0,x_1) = +\infty $.  Again $y_0 \ne y_1$ otherwise the
%horizontal lift of a timelike curve connecting $x_0$ to $x_1$ would
%give a timelike curve connecting $p_0$ to $p_1$. On $M$ there is a
%connecting timelike curve $\sigma(s)$ parametrized with respect to
%proper time whose length satisfies $l(\sigma)> \rho(y_0,y_1)>0$. Let
%$y$ be a maximizing geodesic starting from $y_0$, ending at $y_1$
%and parametrized with length. The curve
%\[
%(\sigma(s),\,  y(\frac{ \rho(y_0,y_1)}{l(\sigma)} s))
%\]
%is timelike and connects $p_0$ with $p_1$, hence a contradiction.
\end{proof}

\begin{theorem} \label{dcon}
If the direct product spacetime $(P,\bm{\tilde{g}})$ is causally
simple then the Lorentzian distance $d: M \times M \to [0,+\infty]$
on the spacetime $(M,g)$ is continuous at the pairs of events
belonging to the set $d^{-1}(C\cup \{+\infty\})\subset M\times M$
where $C=[0,\textrm{diam}(H,h))$ (but it can be continuous in a
larger set).

In particular, if the direct product spacetime $(P,\bm{\tilde{g}})$
is causally simple and $\textrm{diam}(H,h)=+\infty$ then $d: M
\times M \to [0,+\infty]$ is continuous.
\end{theorem}

\begin{proof}It is clear that $d$ is continuous at those points $(x_0,x_1)$
such that $d(x_0,x_1)=+\infty$ because there the Lorentzian distance
is necessarily upper semi-continuous. It is also clear that if
$(x_0,x_1) \notin \bar{J}^{+}$ then $d$ is continuous at $(x_0,x_1)$
because $(\bar{J}^{+})^C$ is an open set where $d$ vanishes. Assume
$d(x_0,x_1)<+\infty$ and $(x_0,x_1) \in \bar{J}^{+}$. From lemma
\ref{uty} and \ref{uty3}

\begin{align}
&{J}^{+} \cap [ \pi^{-1}(x_0)\times \pi^{-1}(x_1)] \nonumber \\
&\quad \subset \{(p_0,p_1): \ p_0=(x_0,y_0), p_1=(x_1,y_1) \textrm{
and
}\rho(y_0,y_1) \le d(x_0,x_1)\} , \label{pis1} \\
&\bar{J}^{+} \cap [ \pi^{-1}(x_0)\times \pi^{-1}(x_1)] \nonumber \\
&\quad =\{(p_0,p_1): \ p_0=(x_0,y_0), p_1=(x_1,y_1) \textrm{ and
}\rho(y_0,y_1) \le D(x_0,x_1)\} \label{pis2},
\end{align}
thus if $d(x_0,x_1)<\textrm{diam}(H,h)$ then $d$ is continuous at
$(x_0,x_1)$ otherwise, $D(x_0,x_1)>d(x_0,x_1)$ and a pair $(y_0,y_1)
\in H \times H$ could be found such that
$d(x_0,x_1)<\rho(y_0,y_1)<\textrm{min}(D(x_0,x_1),\textrm{diam}(H,h))$.
Hence, because of (\ref{pis1}) and (\ref{pis2}), $p_0=(x_0,y_0)$ and
$p_{1}=(x_1,y_1)$, would be such that $(p_0,p_1) \in \bar{J}^+$ but
$(p_0,p_1) \notin J^{+}$ in contradiction with the simple causality
of $P$.

\end{proof}

%
%\[ \bigcap_{i}^{+\infty} \bigcup^{+\infty}_{ k\ge i} M\J^{+}(x_k) \]
%
%
%
%If $y \notin J^{+}(x)$ then since, $M$ is causally simple there is
%an open set $U$, $y \in U$ with $U \cap J^{+}(x) = \emptyset $.
%Thus there is an $N$ such that for $i > N$, $y_i \notin J^{+}(x)$
%or $x \notin  J^{-}(y_i)$. No subsequence of $x_k \to x$ can be
%contained in on $J^{-}(y_i)$ for a certain $i>N$ since otherwise
%$x \in J^{-}(y_i)$. Thus for a certain $\bar{N}_{i}> i$, $x_k
%\notin J^{-}(y_i)$, $k \ge \bar{N}_{i}$, or $y_i \notin
%J^{+}(x_k)$ for $k \ge \bar{N}_{i}>i$.

%Lemmas \ref{csim}, \ref{sim} and \ref{dcon} show that the causal
%simplicity of $(P,\tilde{g})$ implies the causal simplicity of
%$(M,g)$, its maximizing geodesic connectedness and the continuity of
%$d$. Their converse is given by

The next result proves that lemmas \ref{csim}, \ref{sim} and
\ref{dcon} have a converse. Essentially, it proves the equivalence
between the causal simplicity of $(P,\tilde{g})$ and the three
properties of causal simplicity of $(M,g)$, the continuity of $d$ on
$M$ and the maximizing geodesic connectedness of $M$.

\begin{theorem} \label{koi2}
The direct product spacetime $(P,\bm{\tilde{g}})$, $P=M\times H$, is
causally simple if and only if the following three conditions hold
\begin{itemize}
\item[(i)] the spacetime $(M,g)$ is causally simple,
\item[(ii)] the Lorentzian distance $d: M \times M \to [0,+\infty]$ on
the spacetime $(M,g)$ is continuous at (least at) the pairs of
events belonging to the set $d^{-1}(C)\subset M\times M$ where
$C=[0,\textrm{diam}(H,h))$,
\item[(iii)] any pair of causally related events,  $(x_0,x_1)
\in J^{+}$, such that $d(x_0,x_1) <+\infty$ and $d(x_0,x_1) \le
\textrm{diam}(H,h)$ is connected by a maximizing causal geodesic
$\sigma$, $l(\sigma)=d(x_0,x_1)$.
\end{itemize}
In particular if $(M,g)$ is causally simple, maximizing geodesically
connected and $d$ is continuous then $(P,\bm{\tilde{g}})$ is
casually simple whatever the choice of $(H,h)$.

\end{theorem}

\begin{proof} We have only to prove that (i),(ii) and (iii) imply
that $(P,\tilde{g})$ is causally simple. If $(M,g)$ is causally
simple then it is causal and hence $(P,\tilde{g})$ is causal too
(theorem \ref{pcr}). We are going to show that if $(M,g)$ is also
maximizing geodesically connected then points $p_1 \in
\bar{J}^{+}(p_0)\backslash {J}^{+}(p_0)$ do not exist (the past case
is analogous).

Otherwise, by lemmas \ref{bhy} and \ref{uty2}, $x_1 \in I^{+}(x_0)$
and $d(x_0,x_1) \le \rho(y_0,y_1) \le S^+(x_0,x_1)$ where
$p_0=(x_0,y_0)$, $p_1=(x_1,y_1)$. In particular
$0<d(x_0,x_1)<+\infty$ and $d(x_0,x_1) \le\textrm{diam}(H,h)$. There
are two cases, either (a) $d(x_0,x_1) =\textrm{diam}(H,h)$ and thus
$d(x_0,x_1) =\rho(y_0,y_1)$ or (b) $d(x_0,x_1) <\textrm{diam}(H,h)$
and from (ii) $d$ is continuous at $(x_0,x_1)$, that is $
S^+(x_0,x_1)=d(x_0,x_1)$, thus $\rho(y_0,y_1)=d(x_0,x_1)$. In both
cases $\rho(y_0,y_1)=d(x_0,x_1)$. Let $y$ be a minimizing geodesic
starting at $y_0$, ending at $y_1$ and parametrized with respect to
length. Since $d(x_0,x_1)=\rho(y_0,y_1)\le \textrm{diam}(H,h)$ there
is, by assumption (iii) a    maximizing geodesic $\sigma(s)$
connecting $x_0$ to $x_1$ parametrized with respect to proper time.
The curve on $P$
\[
(\sigma(s), y(s))
\]
is lightlike and connects $p_0$ to $p_1$, thus $p_1 \in
{J}^{+}(p_0)$ a contradiction.

%
%Let $q_i=(z_i,w_i) \to p_1=(x_1,y_1)$, $q_i \in J^{+}(p_0)$, by
%lemma \ref{uty}, $\rho(y_0,w_i)=d(x_0,z_i)$
% then the causal curves $\gamma_i$ connecting
%$p_0$ to $q_i$ project into causal curves $\sigma_i$ connecting
%$x_0$ and $z_i$, and since the curves are causal $\rho(y_0,w_i) \le
%l(\sigma_i)$. Since $d$ is continuous
%\[\rho(y_0,y_1)=\limsup_{i \to +\infty} \rho(y_0,w_i) \le \limsup_{i \to
%+\infty} l(\sigma_i) \le \limsup_{i \to +\infty} d(x_0, x_i)=
%d(x_0,x_1), \] hence $q_i$ has a limit $p_1$ in the compact $K$ and
%in particular $p_1 \in J^{+}(p_0)$ thus a contradiction.
%
%
%, and by maximizing geodesic connectedness there would be a
%connecting timelike geodesic $\sigma(s)$ that maximizes the length,
%$l(\sigma)=d(x_0,x_1)$. The curves
%\[
%\gamma_\alpha(s)=(\sigma(s),\, y_0+\alpha s), \qquad \alpha \in
%[-1,1],
%\]
%are causal geodesics (lightlike for $\vert\alpha\vert=1$) of
%endpoint $(x_1, y_0+\alpha d(x_0,x_1))$. The endpoints form a
%compact $K$. Let $q_i=(z_i,w_i) \to p_1=(x_1,y_1)$, $q_i \in
%J^{+}(p_0)$, then the causal curves $\gamma_i$ connecting $p_0$ to
%$q_i$ project into causal curves $\sigma_i$ connecting $x_0$ and
%$z_i$, and since the curves are causal $\vert w_i-y_0\vert \le
%l(\sigma_i)$. Since $d$ is continuous \[\limsup_{i \to +\infty}
%\vert w_i-y_0\vert \le \limsup_{i \to +\infty} l(\sigma_i) \le
%\limsup_{i \to +\infty} d(x_0, x_i)= d(x_0,x_1), \] hence $q_i$ has
%a limit $p_1$ in the compact $K$ and in particular $p_1 \in
%J^{+}(p_0)$ thus a contradiction.
\end{proof}

\begin{remark}
The optimality of the theorem would follow from the independence of
conditions (i),(ii) and (iii). The next three spacetime examples
prove this independence whatever the choice of the constant
$\textrm{diam}(H,h)>0$. Thus they provide three distinct
circumstances for which a direct product spacetime may fail to be
causally simple.

That (i) and (ii) does not imply (iii) is shown by the spacetime of
figure \ref{fig1} where the events $x_0,x_1$, are not connected by a
maximizing geodesic and can be chosen at arbitrary small Lorentzian
distance.

That (ii) and (iii) does not imply (i) can be proved in the
spacetime $(M,g)$, $M=\Lambda\backslash\{o\}$ where $o=(0,0)$ is the
origin of 1+1 Minkowski spacetime $\Lambda$, and the metric is
$g=(t^2+x^2)^{-2}\eta$. Indeed $(M,g)$ is clearly non-causally
simple, and if $(x_0,x_1)$ is such that $J^+(x_0,\Lambda)\cap
J^{-}(x_1,\Lambda)$ contains $o$ then $d(x_0,x_1)=+\infty$ ($d$
distance in $(M,g)$)  thus (ii) and (iii) are satisfied in this
case. If $J^+(x_0,\Lambda)\cap J^{-}(x_1,\Lambda)$ does not contain
$o$ then there are $\bar{x}_0\in I^{-}(x_0,\Lambda)$ $\bar{x}_1\in
I^{+}(x_1,\Lambda)$, such that $(x_0,x_1) \in V=
I^+(\bar{x}_0,\Lambda)\cap I^{-}(\bar{x}_1,\Lambda)$  and $o \notin
V$. But $(V,\eta)$ is globally hyperbolic thus $(V,g)$ is globally
hyperbolic too and has continuous and finite distance function
(because $g$ is conformal to $\eta$) which coincides with the
restriction of the distance function $d$ to $V$ thus $d$ is
continous at $(x_0,x_1)$. Moreover, the global hyperbolicity of
$(V,g)$ implies the existence of a maximizing geodesic connecting
$x_0$ to $x_1$. Thus (ii) and (iii) hold.

It remains to prove that (i) and (iii) does no imply (ii). Take
$M=\Lambda\backslash\{(t,x): x\le 0\}$ and $g=\frac{1}{x} \, \eta $.
The spacetime $(M,g)$ is clearly causally simple. The pairs of
events of the form $(x_0,x_1)$ with $x_0=(b,k-b)$, $x_1=(b,k+b)$
with $b>0$, and $k$ arbitrary constants, have finite Lorentzian
distance which is maximized by a connecting geodesic and which can
be chosen arbitrarily small choosing $b$  sufficiently small. [All
these statements follow from the fact that the maximizing geodesic
can be explicitly calculated. It solves $2\ddot x+\dot x^2-1=0$, and
has equation $x(t)=a+2\ln \cosh(\frac{t-k}{2})$ with
$0<a=b-2\ln\cosh(b/2)$. The fact that the distance goes to zero as
$b$ goes to zero follows from the fact that the length of the
maximizing geodesic is bounded by $\frac{2b}{\sqrt{a}}$ which goes
to  zero.] Nevertheless, there is an infinite discontinuity for $d$
at each pair $(x_0,x_1)$ as above because given $\bar{x}_0\ll x_0$
and $\bar{x}_1 \gg x_1$, it is $d(\bar{x}_0,\bar{x}_1)=+\infty$ as
there are sequences of causal curves connecting $\bar{x}_0$ to
$\bar{x}_1$ which approach a finite vertical segment on the axis
$x=0$. Thus (iii) holds but (ii) does not hold.
\end{remark}

\section{Conclusions}
In this work I studied the correspondence between the causal
properties of $(M,g)$ and those of the warped product $(P,\hat{g})$,
$P=M\times H$, $\hat{g}=g+f^2 h$, with $(H,h)$ a complete Riemannian
manifold. I showed that any statement involving only conformal
properties can be reduced to the case $f=1$, relating $(M,g)$ with
the direct product $(P,\tilde{g})$. An almost complete
correspondence between the causal properties of the two spacetimes
was found, in particular I found a correspondence for the properties
of being distinguishing or non-totally vicious for which no previous
result was available. In the process a formula for the Lorentzian
distance on $(P,\tilde{g})$ in terms of the Lorentzian distance on
$(M,g)$ was obtained.

For causal continuity and causal simplicity the correspondence does
not hold and indeed I gave an explicit counterexample in the latter
case and suggested a possible counterexample for the former case.
Distinct, non conformal invariant, and apparently unrelated
properties must be required on $(M,g)$. The results  which clarify
this issue were theorems \ref{causc}, for causal continuity and
theorem \ref{koi2}, for causal simplicity.

Theorem  \ref{koi2}, obtained here for a spacelike dimensional
reduction geometry (the fibers $\pi^{-1}(x)$ are spacelike), has an
interesting analog in the lightlike dimensional reduction case
\cite{minguzzi06d}. In that work the role of the Lorentzian distance
is replaced by a classical action functional on the base, and the
upper semi-continuity of the Lorentzian distance is replaced by the
lower semi-continuity of the action functional with respect to
endpoints changes.

Finally, some new results on the continuity of the Lorentzian
distance on distinguishing, causally continuous and causally simple
spacetimes were also obtained, see theorem \ref{xdy}, corollary
\ref{fyu} and theorem \ref{pkh}.

%As for the applicability of theorem \ref{koi2} assume you have given
%a spacetime $(M,g)$ and want to prove the  causal simplicity of a
%warped product spacetime $(P,\hat{g})$, with $\hat{g}=g+f(x) \dd
%y^2$. Observe that the causal simplicity of $(P,\hat{g})$ is
%equivalent to the causal simplicity of $(P,\hat{g}/f )$, and
%$\hat{g}/f=q/f+\dd y^2$. Thus, the causal simplicity of
%$(P,\hat{g})$ can be proved checking the three properties of causal
%simplicity, maximizing geodesic connectedness and continuity of the
%distance for the spacetime $(M,g/f)$. Another consequence of the
%theorem is that any spacetime $P$ with metric of the form $g+\dd
%y^2$, if causally simple, has immediately other properties, as the
%continuity of the Lorentzian distance and the connectedness through
%maximizing geodesics.

\section*{Acknowledgements}
I warmly thank M. S\'anchez, this work, in its early stage of
development, has benefited from his suggestions especially in
connection with theorem \ref{pkh} and lemma \ref{cxz}.

%\bibliography{../../bibliografie/simultaneity,../../bibliografie/libri,../../bibliografie/miei,../../bibliografie/mieiPreprints,../../bibliografie/mieiProceedings}
%\bibliographystyle{plain}

\end{document}